\documentclass[epj]{svjour}
\usepackage{graphicx}
\def\makeheadbox{{%  remove EPJC header
}}
\begin{document}
\pagestyle{empty}
\title{The dark connection between the Canis Major dwarf, the Monoceros ring, the gas flaring,
the rotation curve and the EGRET excess of diffuse Galactic Gamma Rays}
\author{W. de Boer, I. Gebauer, M. Weber, C. Sander, V. Zhukov\inst{1}
\thanks{\emph{Email:} wim.de.boer@cern.ch} %
\and D. Kazakov\inst{2}% etc
}                     % Do not remove
\institute{Institut f\"ur Experimentelle Kernphysik, University of Karlsruhe,
Postfach 6980, 76128 Karlsruhe, Germany \and Bogoliubov Laboratory of Theoretical
Physics, Joint Institute for Nuclear Research,
           141980 Dubna, Moscow Region, Russia}
%
%needed for epjc \date{Received: date / Revised version: date}
\date{}
\abstract{ The excess of diffuse galactic gamma rays above 1 GeV, as observed by
the EGRET telescope on the NASA Compton Gamma Ray Observatory, shows all the key
features from Dark Matter (DM) annihilation: (i) the energy spectrum of the excess
is the same in all sky directions
   and is consistent with the gamma rays expected for the  annihilation
   of  WIMPs with a  mass   between 50-100 GeV;
(ii) the intensity distribution of the excess in the sky is used to
   determine the halo profile, which was found to correspond to the
   usual profile from N-body simulations with additional substructure
   in the form of two doughnut-shaped  structures  at radii of 4 and 13 kpc;
(iii)  recent N-body simulations of the tidal disruption of the Canis Major
    dwarf galaxy show that it is a perfect progenitor of the ringlike
    Monoceros tidal stream of stars at 13 kpc with ring parameters in
     agreement with the EGRET data;
(iiii)  the mass of the outer ring is so large, that its gravitational effects
    influence both the gas flaring and the rotation curve of the Milky Way.
    Both effects are clearly observed in agreement with the DMA interpretation
    of the EGRET excess.
%
%
%\PACS{
%      {Gamma Rays}{ observations, theory, positron annihilation} \and
%      {Milky Way} {dark matter halo, cosmic rays, propagation, structure,  rotation curve,
%dwarf galaxies} \and {Cosmology} {dark matter annihilation, dwarf galaxies,
%structure formation} \and {Elementary Particles} {dark matter annihilation,
%neutralinos, Supersymmetry} }
} %end of abstract
\maketitle
\section{Introduction}
 If dark matter  (DM) is
created thermally during the Big Bang the present relic density  is inversely
proportional to $\langle\sigma v\rangle$, the annihilation cross section $\sigma$
of DM particles, usually called WIMPS (Weakly Interacting Massive Particles), times
their relative velocity. The average is taken over these velocities. This inverse
proportionality is obvious, if one considers that a higher annihilation rate, given
by $\langle\sigma v\rangle n_\chi$, would have reduced the relic density before
freeze-out, i.e. the time, when the expansion rate of the Universe, given by the
Hubble constant, became equal to or larger than the annihilation rate.  For the
present value of $\Omega h^2=0.105 \pm 0.008$, as measured by WMAP \cite{wmap}, the
thermally averaged total cross section at the freeze-out temperature of $m_\chi/22$
must have been around $3\cdot 10^{-26} ~{\rm cm^3s^{-1}}$ \cite{jungman}. If the
s-wave annihilation is dominant, as expected  in many supersymmetric models, then
the annihilation cross section is energy independent, i.e. the cross section given
above is also valid for the cold temperatures of the present universe \cite{susy}.
Such a large cross section will lead to a production rate of mono-energetic
quarks\footnote{The quarks are mono-energetic, since the kinetic energy of the cold
dark matter particles is expected to be negligible with the mass of the particles,
so the energy of the quarks equals the mass of the WIMP.} in our Galaxy, which is
40 orders of magnitude above the rate produced at any accelerator. The
fragmentation of these  quarks will lead to a large flux of gamma rays with a
characteristic energy spectrum quite different from the background of cosmic ray
interactions with the interstellar material of the Galaxy. In addition, gamma rays
have the advantage that they point back to the source and do not suffer energy
losses, so they are the ideal candidates to trace the dark matter density. The
charged components interact with Galactic matter and are deflected by the Galactic
magnetic fields, so they do not point back to the source. Therefore the charged
particle fluxes have large uncertainties from the pro\-pagation models, which
determine how many of the produced particles arrive at the detector. For gamma rays
the propagation is straightforward: only the ones pointing towards the detector
will be observed.

An excess of diffuse gamma rays compatible with dark matter annihilation (DMA) has
indeed been observed by the EGRET telescope on board of NASA's CGRO (Compton Gamma
Ray Observatory)\cite{us}. The excess was observed in all sky directions, which
would imply that DM is not dark anymore, but shining in gamma rays.  Of course,
such an important observation needs to be scrutinized heavily. Before discussing
the criticism the evidence and new confirmation from N-body simulations and the gas
flaring is presented in the next section.
\begin{figure}
\begin{center}
\vspace*{-0.65cm}
 \includegraphics [width=0.42\textwidth,clip]{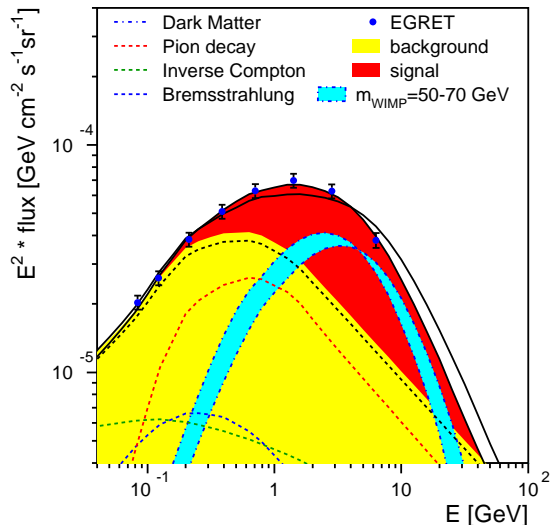}
 \caption[]{Fit of the shapes of background and
    DMA signal to the EGRET data in the direction of the Galactic centre
     The light shaded (yellow)
    area indicates the background using the shapes known from accelerator experiments,
     while the dark shaded (red)
    area corresponds to the signal contribution from DMA for a 60 GeV WIMP mass.
    The intermediate (blue) shaded area corresponds to a variation of the WIMP mass
    between 50 and 70 GeV.}
 \label{fig2}
\end{center}
\end{figure}

\section{The DMA interpretation of the EGRET excess of diffuse Galactic gamma
rays}\label{egret}

 The EGRET excess on diffuse gamma
rays was first observed by Hunter et al. in 1997 \cite{hunter}. Below 1 GeV the CR
interactions describe the data perfectly well, but above 1 GeV the data are up to a
factor two above the expected background. The excess shows all the features of DMA
annihilation for a WIMP mass between 50 and 70 GeV \cite{us}. Especially, the two
basic constraints expected from any indirect DMA signal are fulfilled: (i) the
excess should have the same {\it spectral shape} in all sky directions. (ii) the
excess should be observable in a large fraction of the sky with an {\it intensity
distribution} corresponding to the gravitational potential of our Galaxy. The
latter means that one should be able to relate the distribution of the excess to
the rotation curve. Both conditions, which form a formidable constraint, are met by
the EGRET data \cite{us}. In addition, the results are perfectly consistent with
the expectations from Supersymmetry\cite{susy,sander}.

The analysis of the EGRET data is simplified by the fact that the spectral shapes
of the DMA contribution and the background from CR interactions with the gas of the
disk are well known from accelerator experiments: (i) the DMA signal should have
the gamma ray spectrum from the fragmentation of mono-energetic quarks, which has
been studied in great detail at LEP. (ii) the background in the energy range of
interest is dominated by cosmic ray (CR) protons hitting the hydrogen of the disk.
Therefore the dominant background spectral shape is known from fixed-target
experiments. Given that these shapes are known from the two best studied reactions
in accelerator experiments allows to fit these {\it known} shapes to the observed
gamma ray spectrum in a given sky direction and obtain from the fitted
normalization constants the contribution of both, background and annihilation
signal. So in this case one does not need propagation models to estimate the
background, since the data itself calibrates the amount of background. In addition,
fitting the shapes eliminates the uncertainties from the overall normalization
error in the data and only the reduced point-to-point errors have to be taken into
account. A typical gamma ray energy spectrum is shown in Fig. \ref{fig2}, which
clearly shows the rather distinct spectral shapes for DMA and background, so the
two normalization constants of DMA and background are not strongly correlated. The
blue area corresponds to the difference in WIMP mass between 50 and 70 GeV. The
latter gives already a considerably worse fit, since the total flux is too low near
the maximum and too high for the highest bin, so the range 50-70 GeV is the
preferred WIMP mass.
\begin{figure}
\begin{center}
\vspace*{-0.8cm}
 \includegraphics [width=0.5\textwidth,clip]{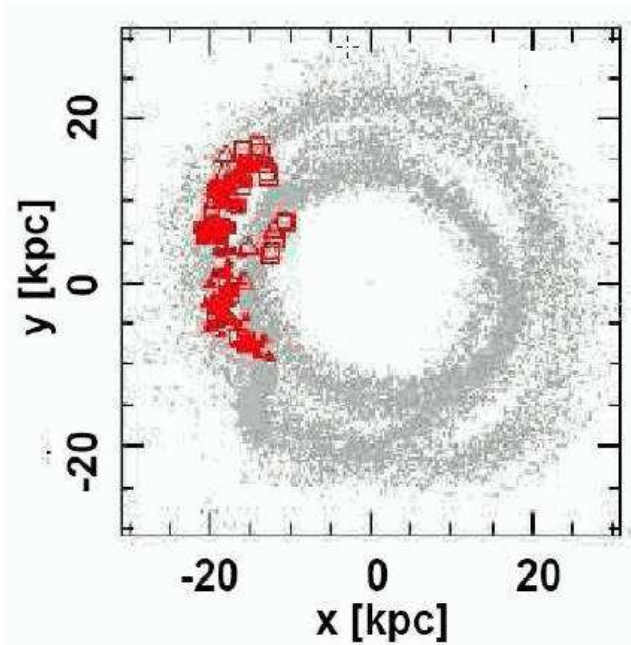}
 \caption[]{Results of an N-body simulation of the tidal disruption of the
 Canis Major dwarf Galaxy, whose orbit was fitted to the observed stars (red
 points). The simulation predicts a ringlike structure of dark matter with a
 radius of 13 kpc. From \cite{penarrubia}.
    }
 \label{fig3}
\end{center}
\end{figure}
The DM halo, as determined from the dark matter normalization factors in 180
independent sky directions, shows  doughnut-like structures at 4 and 13 kpc
\cite{us}. More details can be found in the contribution to these
proceedings\cite{weber}. The ring at 4 kpc (inner ring) coincides with the ring of
dust in this region. The dust is presumably kept there because of a gravitational
potential well, which is provided by the ring of DM. The ring at 13 kpc (outer
ring) is thought to originate from the tidal disruption of the Canis Major dwarf
galaxy, which circles the Galaxy in an almost circular orbit coplanar with the disk
\cite{penarrubia1,ibata1}. Three independent observations confirm this picture of
the ring originating from the tidal disruption of a dwarf galaxy:
\begin{figure}
\begin{center}
 \includegraphics [width=0.5\textwidth,clip]{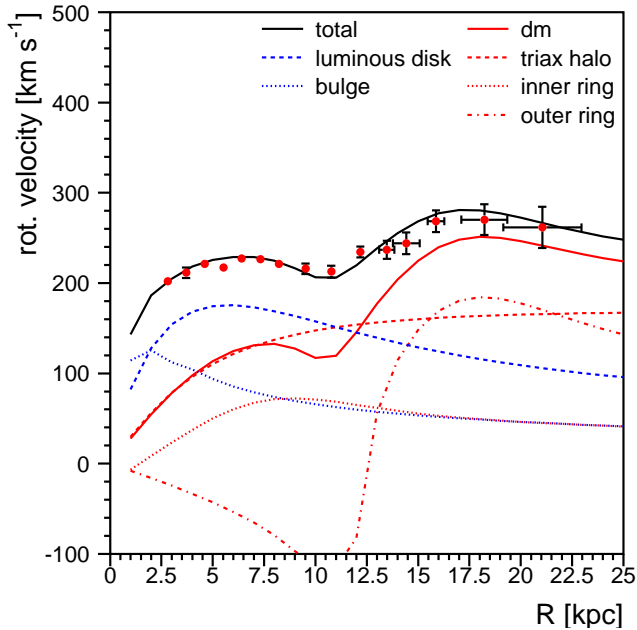}
 \caption[]{The rotation curve with the contributions of the bulge, the disk, the triaxial dark matter
 halo and the two ringlike structures. The outer ring causes the peculiar change of slope in the
 rotation curve at about 11 kpc. From  \cite{us}.}
 \label{fig4}
\end{center}
\end{figure}

(i)a ring of DM is expected in this region from the observed ring of stars, called
Monoceros ring, which was discovered first with SDSS data \cite{newberg,yanny}.
Follow-up observations \cite{ibata} found that this structure surrounds the
Galactic disk as a giant ring (observed over  100 degrees in latitude) at
Galactocentric distances from ~ 8 kpc to ~ 20 kpc. Tracing this structure with
2MASS M giant stars, \cite{rocha-pinto} suggested that this structure might result
from the tidal disruption of a merging dwarf galaxy. N-body simulations show indeed
that the overdensity in Canis Major is indeed a perfect progenitor for the
Monoceros stream  and they predict  a DM ring at 13 kpc with a low ellipticity and
almost coplanar with the disk, as shown in Fig. \ref{fig3} (from Ref.
\cite{penarrubia}). The orientation of the major axis at an angle of 20 degrees
with respect to the axis sun-Galactic centre and the ratio of minor to major axis
around 0.8 agree with the EGRET ring parameters given in Ref. \cite{us}. This
correlation with the EGRET excess lends both support to the DMA interpretation of
the EGRET excess, the connection between the Monoceros stream and the Canis Major
dwarf galaxy and its outer galactic origin. The overdensity of stars forming the
Canis Major dwarf is sometimes defended as being a warp of the Galactic disk (see
discussions e.g. in \cite{ibata2}).

(ii) Such a massive ring structure influences the rotation curve in a peculiar way:
it decreases the rotation curve at radii inside the ring and increases it outside.
This is apparent from the change in direction of the gravitational force from the
ring on a tracer, since this force decreases the force from the galactic centre for
a tracer inside the ring, but increases it outside the ring. This is indeed
observed as shown in Fig. \ref{fig4}, where the negative contribution of the outer
ring is clearly visible.

\begin{figure}
\begin{center}
 \includegraphics [width=0.5\textwidth,clip]{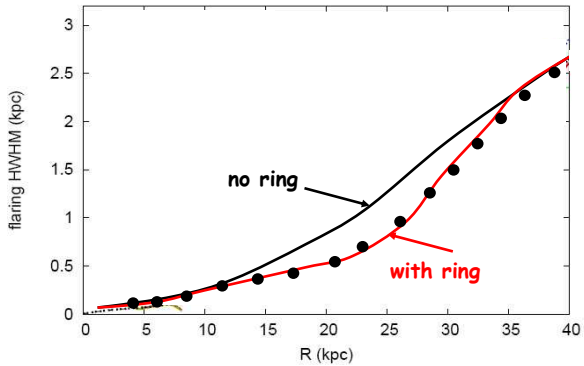}
 \caption[]{The half-width-half-maximum (HWHM) of the gas layer of the Galactic Disk as function
 of the distance from the Galactic center. Clearly, the fit including a ring of dark matter
 above 10 kpc describes the data much better. Adapted from data in \cite{kalberla}.  }
 \label{fig5}
\end{center}
\end{figure}

(iii) A direct proof of the large amount of DM mass in the outer ring comes from a
recent analysis of the gas flaring in our Galaxy \cite{kalberla}. Using the new
data of the LAB survey of the 21 cm line in our Galaxy led to a precise measurement
of the gas layer thickness up to radii of 40 kpc. The increase of the half width of
the layer after a decrease to half its maximum value (HWHM)  as function of
distance is governed by the decrease in gravitational potential of the disk. The
outer ring increases the gravitational potential above 10 kpc, which is expected to
reduce the gas flaring. Only after taking the ring like structure into account the
reduced gas flaring in this region could be understood. The effect is shown in Fig.
\ref{fig5}. A fit averaged over all longitudes requires a DM ring with a mass of
$2.10^{10}$ solar masses, in rough agreement with the EGRET excess.

Clearly, these three independent astronomical observations need  a  ringlike DM
structure above 10 kpc, thus  providing independent evidence for the DMA
interpretation of the EGRET excess.
\section{Alternatives to the DMA interpretation of the EGRET excess and criticism}
Alternative models for the EGRET excess without DM  have to assume that the locally
measured fluxes of protons and electrons are not representative for our galaxy
\cite{optimized}. In this case the cosmic ray spectrum of protons and electrons is
not taken to be the locally observed one, but modified to increase the gamma ray
spectrum at high energies. This requires a strong break in the injection spectrum
of electrons and protons in order not to change the gamma ray spectrum below 1 GeV,
but only above 1 GeV. Such a change in the shape of the spectrum is unexpected,
since the fast diffusion as compared to the energy loss time equalizes the spectrum
everywhere in the Galaxy, in agreement with the fact that the spectrum in all
directions is observed to be the same.

Another  explanation is provided by tuning the efficiency of the EGRET spectrometer
to simulate DMA \cite{stecker}. However, this requires the efficiency already to be
modified around 1 GeV and reaching a change in efficiency of 100\% at 10 GeV in
clear disagreement with the calibration error in a photon beam before
launch\cite{egret_cal} and the residual uncertainties below 20\% during the flight
after correcting for time dependent effects\cite{egret_cal1}. Although there is
some uncertainty in the efficiency of the veto counter at higher energies because
of the backsplash from the calorimeter, this effect should not start at 1 GeV. Even
if the errors are larger than estimated, it would be a remarkable coincidence  that
the excess corresponds exactly to the very specific sharply falling spectrum from
the fragmentation of mono-energetic quarks! Furthermore, if one only fits the shape
of the gamma ray spectrum all common normalization errors cancel, so one is only
sensitive to the relative point-to-point errors in the data, which are considerably
smaller.

 Among the most important criticism was a paper by \cite{bergstrom1}
claiming that the antiproton flux from DMA, using the DM distribution from the
EGRET excess, would be an order of magnitude higher than the observed antiproton
flux. They used a simple propagation model assuming the propagation of charged
particles to be the same in the halo and the disk. Such common propagation models
describe all present data, but this is {\it no} guarantee that they represent the
truth, since the data is observed locally and inferring the propagation in the
whole Galaxy from local data does not have a unique solution. E.g. Galactic wind
models have anisotropic propagation, not only because of the convection, but also
because the diffusion in the halo is expected to be much faster than in the disk
\cite{breitschwerdt_gamma}.

Such an anisotropic diffusion model leads to a strong reduction of the antiproton
flux from DMA, since the antiprotons are preferentially produced in the halo and
stay there or move to outer space instead of diffusing randomly. Therefore, instead
of rejecting the DMA interpretation of the EGRET excess by models based on
isotropic diffusion without regular magnetic fields, DMA can be used as an
opportunity to construct an anisotropic propagation model. It is shown in a
separate contribution to this conference that such a model can simultaneously
describe the EGRET excess, the antiproton flux and the production of secondaries,
if DMA is introduced as a source term in the model \cite{gebauer}.

In summary, the gamma rays play  a very special role for indirect DM detection,
since they point back to the source. Therefore the gamma rays provide a perfect
means to reconstruct the dark matter halo by observing the intensity of the gamma
ray emissions in the various sky directions. The halo profile can in turn be used
to calculate the shape of the rotation curve. Fitting the background and DMA shapes
to the data in 180 independent sky directions reveals indeed a WIMP mass around 60
GeV in all sky directions and a halo profile with substructure consistent with the
rotation curve, N-body simulations and the gas flaring. For the halo profile one is
only interested in the relative contributions in the various sky directions, so
here all experimental errors largely cancel, since the EGRET satellite does not
care in which direction it measures. 

Therefore we consider DMA is a viable explanation of the EGRET excess of diffuse
Galactic gamma rays, especially since it is observed with the same shape of the
fragmentation of mono-energetic quarks in all sky directions and the intensity
distribution of the excess traces the DM profile, as shown independently by the
rotation curve, the gas flaring and the N-body simulation of the disruption of the
Canis-Major satellite galaxy.
%\section{Acknowledgements}
% I thank  my close collaborators I. Gebauer, A. Gladyshev, D.
%Kazakov, C. Sander, M. Weber and V. Zhukov for their contributions
%to this interesting project.

This work was supported by the DLR (Deutsches Zentrum f\"ur Luft- und Raumfahrt)
 and a grant from the
DFG (Deutsche Forschungsgemeinschaft, Grant 436 RUS 113/626/0-1).
%\section{References}

\end{document}